\documentclass[twocolumn,prb,aps,nofootinbib,amsmath,amssymb]{revtex4}
\usepackage[dvips]{graphicx}
\usepackage{array}

\newcommand {\be}{\begin{equation}}
\newcommand {\ee}{\end{equation}}

\begin{document}

\title{Modeling Diffusive Dynamics in Adaptive Resolution Simulation of Liquid Water }
\author{Silvina Matysiak}
\author{Cecilia Clementi}
\affiliation{Department of Chemistry, Rice University, 6100 Main Street, Houston, Texas 77005}
\author{Matej Praprotnik}
\altaffiliation{On leave from the National Institute of Chemistry, Hajdrihova 19,
                 SI-1001 Ljubljana, Slovenia.}
\author{Kurt Kremer}
\author{Luigi Delle Site}
\affiliation{Max-Planck-Institut f\"ur Polymerforschung, Ackermannweg 10, D-55128 Mainz, Germany}


\date\today
\begin{abstract}
 We present a dual-resolution molecular dynamics (MD) simulation of liquid water employing a
recently introduced Adaptive Resolution Scheme (AdResS). 
The spatially adaptive molecular resolution procedure allows 
for changing from a coarse-grained to an all-atom representation and
vice-versa on-the-fly. In order to find the most appropriate
  coarse-grained water model to be employed with AdResS 
we first study the accuracy of different coarse-grained water models 
in reproducing the structural properties of the all-atom system. 
Typically, coarse-grained molecular models
have a higher diffusion constant than the corresponding all-atom
models due to the reduction in degrees of
freedom (DOFs) upon coarse-graining that eliminates the
fluctuating forces associated with those integrated-out molecular
DOFs. Here, we introduce the methodology
to obtain the same diffusional dynamics across different resolutions. 
We show that this approach leads to the correct description of 
essential thermodynamic, structural and dynamical properties of liquid
water at ambient conditions.
\end{abstract}

\maketitle
\section{Introduction }
Water is a typical example where the interplay between different length scales determines the relevant
properties of the system. Including explicit water in large biomolecular simulations is crucial \cite{Pal:2002, Cheung:2002} but normally not feasible due to the large number of water molecules needed to describe biomolecular function. The computational requirements of
all-atom simulations in explicit water usually do not allow to investigate biologically relevant time-scales\cite{Praprotnik:2005:2}.
Most computational approaches renormalize the role of water into the definition of
effective inter-residue interactions \cite{Cheung:2002} or as a continuous field, in order to focus
the computational efforts to simulate the biomolecule. However, the discrete nature of water
molecules can play a role in, e.g., controlling protein
functionality or the case of interactions between hydrocolloids and water, where
hydrogen bonding is important, etc.
To overcome this problem, a multiscale modeling of water can be very advantageous to speed up the simulation without losing physics-chemical accuracy. 
 
Multiscale modeling is emerging as a viable way 
to bridge the various length and time scales involved in complex molecular
systems\cite{Broughton:1999,Csanyi:2004,Kremer:2004,Koumoutsakos:2005,
Neri:2005,Backer:2005,Abrams:2005,Fabritiis:2006,Lyman:2006,
Praprotnik:2007, Praprotnik:2007:2}. 
In general, to build a multiscale model two main issues need to be addressed.
The first issue consists in mapping the system into a robust reduced model
while preserving the here relevant physico-chemical properties, 
i.e., radial distribution functions, pressure, and temperature, of the
reference all-atom system.
The second issue involves the definition
of a robust and physically accurate procedure to smoothly join the different resolutions.
We have recently proposed the Adaptive Resolution Scheme
  (AdResS)\cite{Praprotnik, Praprotnik1} 
to address both issues for a system of water molecules \cite{water_jphys}.
Typically, the reduction in the number of the system degrees of freedom (DOFs) upon coarse-graining
introduces a time scale difference between the coarse-grained and explicit region.
Although in certain instances the difference in time scale may be
advantageous for reaching longer simulation times, in other cases having the correct diffusional dynamics is crucial. This is the case, for instance, if a dynamical property is focus of an investigation as in the translocation of biopolymers through membrane nanopores \cite{Matysiak_PRL}.\\

\begin{figure}[ht!]
\centering
\includegraphics[width=0.8\linewidth,clip=]{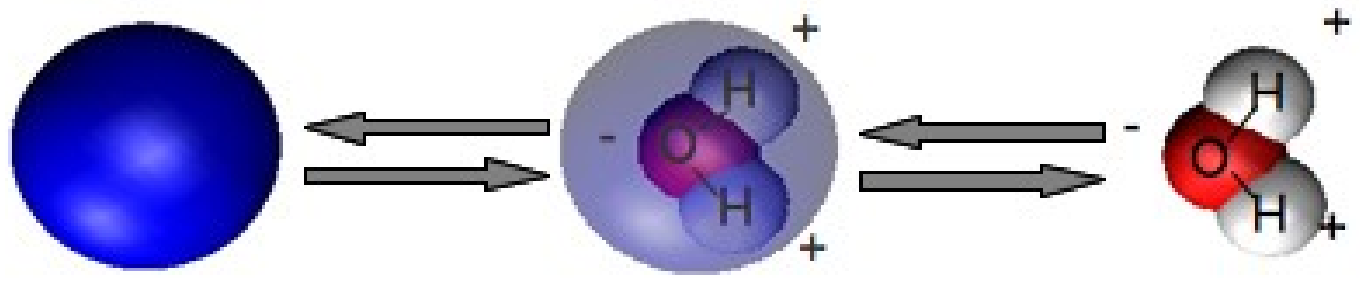}
\includegraphics[height=0.8\linewidth,clip=]{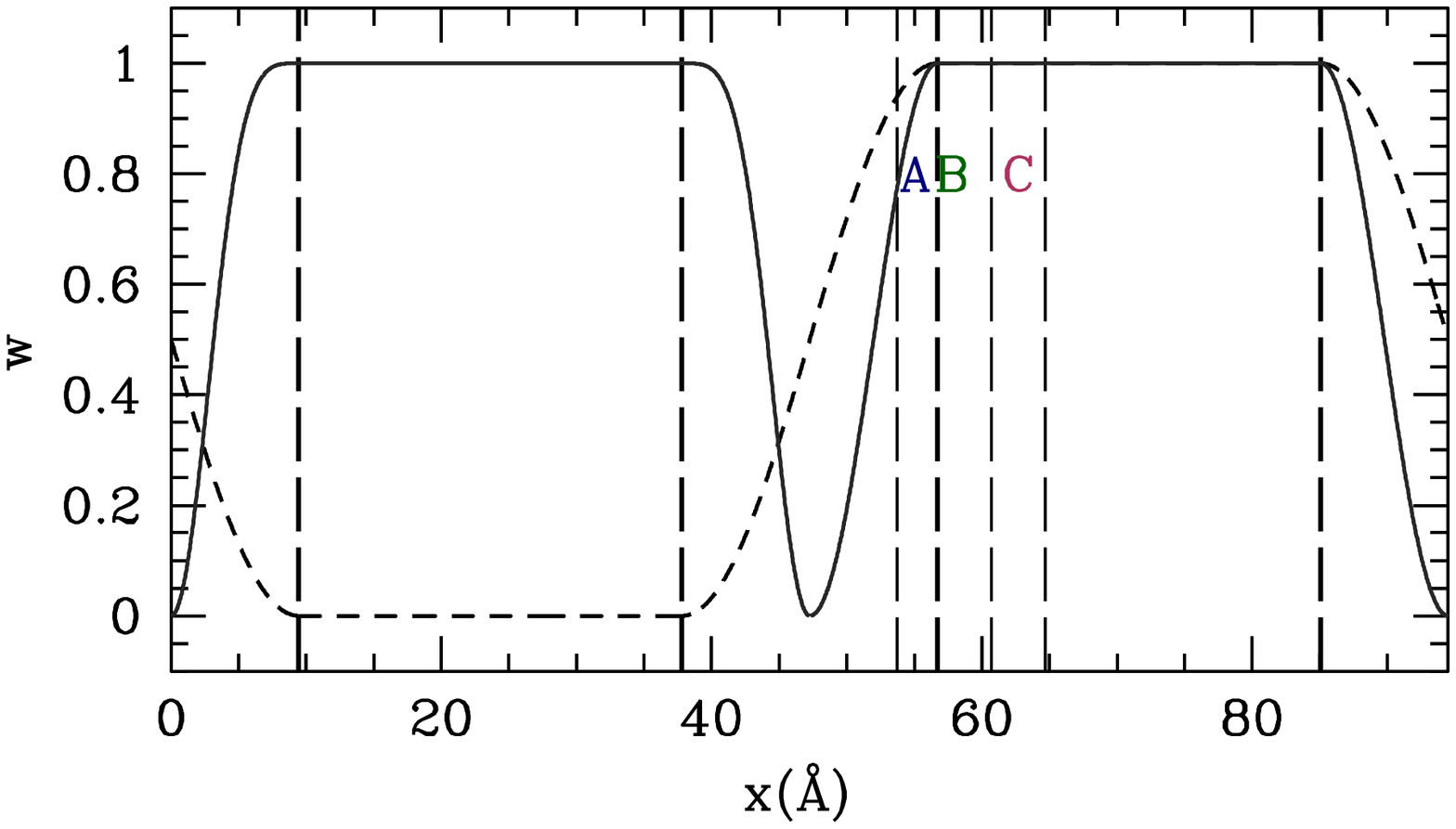}
\vspace{-3cm}
\caption{\scriptsize (Color online) Top figure: The coarse-grained molecule is represented at the left, and the
all-atom water molecule is represented at the right. The middle hybrid molecule interpolates
between the two. 
Bottom figure: The dashed line represents the weighting function w(x) $\in [0,1]$ defined in Ref. \cite{Praprotnik} and discussed in the text. The
values w=1 and w=0 correspond to the atomistic and coarse-grained regions of the hybrid 
atomistic/mesoscopic system. 
The full line represents the interface correction weighting function s(w) defined in Figure \ref{s(w)}. The value s=1 corresponds to the atomistic and coarse-grained regions, while s=0 when w=1/2.
The explicit and interface regions are divided in three regions(A,B,C) of width 4 \AA \; for computing the cosine distribution in Figure \ref{fig_cosine}. 
\label{fig.system} }
\end{figure}

In this paper, in order to justify our choice of the recently
  introduced single-site water model\cite{water_jphys} as the optimal coarse-grained water
  model to be used with AdResS (when using  the standard three-site
  TIP3P water model\cite{Jorgensen} as the all-atom water model), 
we present firstly a detailed analysis and comparison of 
newly developed coarse-grained water models having a different number
of DOFs (from $3$ to $9$) with the single-site water model\cite{water_jphys}
as well as the TIP3P water model\cite{Jorgensen}.
In this way we can study the contribution of separate DOFs
of the all-atom water model to the definition of its structural
  properties. As it turns out the single-site
  water model can reproduce the structural and thermodynamical
  properties of the all-atom system that are relevant for the multiresolution simulation
  equally well as the more sophisticated models. Therefore, we have chosen
  it as the coarse-grained model in our adaptive resolution simulations. 
Next, we introduce the methodology to obtain the same diffusional dynamics across the different
resolutions in the hybrid atomistic/coarse-grained (ex-cg) model
system  composed of explicit and coarse-grained molecules as presented in Figure \ref{fig.system}.

\section{Mapping of a coarse-grained model to the all-atom representation}
In order to map a coarse-grained to an all-atom model, we have
to construct an effective potential
between coarse-grained molecules in a way that the atomistic structural properties
are reproduced.
We numerically build such effective Hamiltonian following an "inverse statistical-mechanics" approach \cite{Lyubarstev,Matysiak_JMB2004,Matysiak_JMB2006}.
In order not to bias a priori the potential function with the choice of a particular
functional form, we introduce a grid approximation, expressing a general
pairwise Hamiltonian in the form:
\be
H=\sum_{\alpha}U_{\alpha}S_{\alpha}
\ee
where $S_{\alpha}$ is the number of coarse-grained particles in the interval
$\in (r_{\alpha}, r_{\alpha}+dr)$ and $U_{\alpha}$ is the discretized value of
the potential in the interval $(r_{\alpha}, r_{\alpha}+dr)$.
The averages $\langle S_{\alpha} \rangle$, which can be interpreted as a discretized
radial distribution function (rdf), are functions of the set of parameters $\{U_{\alpha}\}$.
Values of the potential parameters
$\{U_{\alpha}\}$ that reproduce the rdf of the all-atom model 
can be obtained iteratively as follows:
\begin{itemize}
\item The center-of-mass rdf $g(r)$ is computed from all-atom simulations and used
as a target function. The potential of mean forces is assumed as an initial
approximation to the effective potential function:
\be
U_{\alpha}^{0}= -kT \ln g(r_{\alpha})
\ee
where $U_{\alpha}$ is the potential at a distance $r_{\alpha}$.
\item By comparing the atomistic center-of-mass rdf with the rdf
obtained for the coarse-grained model at the n-th iteration (with corresponding
potential parameters $U_{\alpha}^{n}$) the $(n+1)$-th correction to the
effective potential can be found. Namely, the correction to the potential
parameters $U^{n}_{\alpha}$ is given by the solution of the system of linear
equations \cite{Lyubarstev}:
\be
\Delta\langle {S_{\alpha}}\rangle=\sum_{\gamma}{\frac{\partial\langle
S_{\alpha}\rangle}{\partial U_{\gamma}} \Delta U_{\gamma}}
\label{eq}
\ee
where:
\be
\frac{\partial\langle S_{\alpha}\rangle}{\partial U_{\gamma}}=-\beta
\biggl(\langle S_{\alpha}S{\gamma}\rangle-\langle S_{\alpha} \rangle \langle
S_{\gamma} \rangle \biggr)
\label{system}
\ee
\item Ensemble averages $\langle S_{\alpha}S_{\gamma}\rangle$ and $\langle
S_{\alpha}\rangle$,$\langle S_{\gamma}\rangle$ are computed from MD simulations
(with potential parameters $U_{\alpha}^{n}$), and the correction $\Delta U_{\alpha}^{n}$
to the potential $U_{\alpha}^{n}$ is obtained by solving the system of linear equations (\ref{eq}):
\be
U_{\alpha}^{n+1}=U_{\alpha}^{n}+\Delta U_{\alpha}^{n}.
\ee
\item The corrected potential parameters $U_{\alpha}^{n+1}$ are then used to
perform molecular dynamics simulation with the coarse-grained model and calculate new ensemble averages for the quantities
$\langle S_{\alpha}S_{\gamma}\rangle$ and
$\langle S_{\alpha}\rangle$,$\langle S_{\gamma}\rangle$.
\end{itemize}
The points above are repeated to determine a new set of corrections
$\Delta U_{\alpha}^{n+1}$.
The procedure is repeated until convergence is reached within the statistical simulation error.\\
Additionally, to match the pressure of the coarse-grained to the
all-atom model, after each iteration, a weak constant force is
added to the effective force in such a way that the total
effective force and potential energy are zero at the fixed cutoff
distance $r_{c}$\cite{Reith, Praprotnik1}.
\be
\Delta U(r)=U_{o}\biggl(1-\frac{r}{r_{c}}\biggr)
\ee
We use a value of $r_{c}=7$ \AA \; as the rdf $\approx 1$ for $r>r_{c}$.
Depending on the pressure in the current iteration being above or below the target
value corresponding to the pressure of the reference all-atom system, $U_{o}$ can
assume a positive or negative value.\\
For the atomistic representation of water we selected the rigid TIP3P model \cite{Jorgensen} 
that gives a reasonable description of the behavior of liquid water around
standard (physiological) temperature and pressure, and is not too expensive computationally.
The potential function of rigid TIP3P involves a rigid water with three
interaction sites that coincide with the atomic positions. 
The model uses atom-centered point charges to represent the
electrostatic, i.e., positive charges on the hydrogens
and a negative charge on oxygen ($e_{O}=-2e_{H}$). The van der Waals
interaction between two water molecules is modeled by the
Lennard-Jones potential with just a single interaction site per
molecule centred on the oxygen atom; the van der Waals interactions
involving the hydrogens are omitted. This gives the following potential:
\be
H=\sum_{i}\sum_{j}\frac{e_{i}e_{j}}{r_{ij}}+\frac{A}{r_{OO}^{12}}-\frac{C}{r_{OO}^6}
\label{pot}
\ee
The values of the parameters $e, A,$ and $C$ are assigned to reproduce reasonable structural and energetics properties of liquid water \cite{Jorgensen}.


\section{Coarse-grained models for water}
\begin{figure}[ht!]
\centering
\includegraphics[height=0.7\linewidth,clip=]{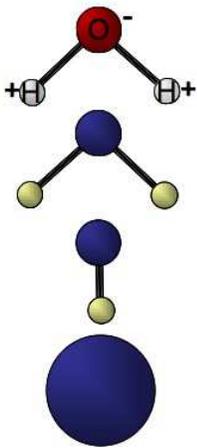}
\caption{\scriptsize (Color online) Cartoon of different water models
  employed in this study. From top to bottom: The rigid TIP3P water
  model, and three-site, two-site, one-site coarse-grained water models, respectively.}\label{water_models}
\end{figure}

Many simplified coarse-grained models have been developed to
reproduce qualitatively the structural properties of
water\cite{Dill}. Existing coarse-grained
models can of course only reproduce thermodynamical properties of all-atom
  water to a certain extend\cite{Ashbaugh,Vothjcp2005,Gordon,Soper,Nezbeda}. 
By using the procedure described above, we have designed water models
at different level of coarse-graining (three-site, two-site, and
one-site) schematically depicted in Figure \ref{water_models}.
The gradually increasing level of coarse-graining allows us then to
study the contribution of different DOFs to the
  structural and thermodynamical properties of the
reference all-atom system. In this way we can determine the most
  appropriate coarse-grained model to be used in our adaptive
  resolution simulations.

\subsection*{Three-site interaction model}
\begin{figure}[h]
\centering
\includegraphics[height=\linewidth,clip=]{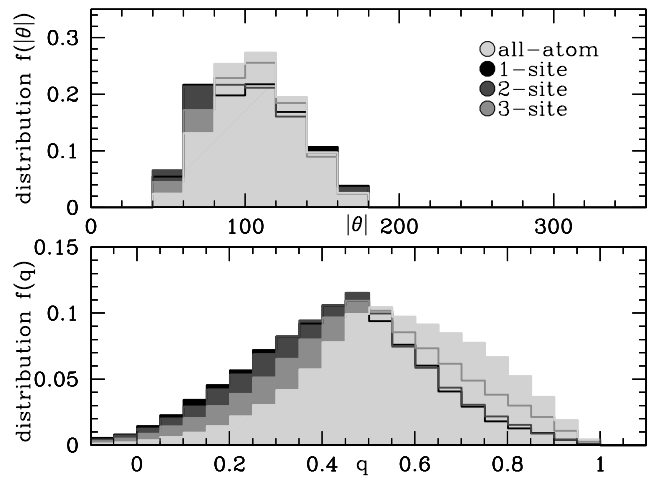}
\caption{\scriptsize Top figure: The center-of-mass angular distribution between three nearest
neighbors for all-atom, 1-site coarse-grained, 2-site  coarse-grained and 3-site coarse-grained models for water.
Bottom figure: The analogous distributions of the orientational order parameter $q$. 
The center-of-mass rdf of the different water models coincides exactly with the all-atom rdf (not shown). 
\label{fig.struct} }
\end{figure}

A first level of coarse-graining can be introduced 
by preserving the atomic positions of each atom in the
TIP3P water model and replacing the explicit electrostatic
interactions (the first term in Eq. (\ref{pot})) with effective
short-ranged interactions.
This model preserves explicitly the H-bond directionality but it is computationally less expensive than TIP3P because it removes the long-ranged Coulombic term.
The calculated all-atom site-site rdfs are used as input to build the OO, OH and HH effective potentials (see Eqns. \ref{system}).
In order to quantify the agreement of the rdfs (corresponding to the all-atom and coarse-grained model)
we introduce the penalty function $f_{p}$ \cite{Praprotnik} defined as
\be
f_{p}=\int \biggl[g^{cg}(r/\sigma_{OO})-g^{ex}(r/\sigma_{OO})\biggr]^{2}exp(-r/\sigma_{OO})d(r/\sigma_{OO}),
\ee
where $g^{ex}$ and $g^{cg}$ are the reference site-site rdf of the atomistic and coarse-grained
system, respectively, and $\sigma_{OO}$ is the Lennard-Jones constant
of the $TIP3P$ water model\cite{Jorgensen}.
The extremely low values of the $f_{p}$ obtained ($f_{p}$ of the OO rdf is $7.9*10^{-5}$, for OH is $8.2*10^{-5}$
and HH is $1.1*10^{-4}$) for the optimized effective potential for the three-site 
model indicates a perfect matching of the rdfs. The optimized
  effective potentials are as expected very similar to the
  effective potentials of other previously introduced 3-site coarse-grained water models\cite{Soper,Lyubarstev1}.  
In addition, to
check the angular properties (that are not completely defined
by the rdf) we computed the distribution of the angle $\theta$ formed between the
center-of-mass of three nearest neighbor molecules (Figure \ref{fig.struct}, top), and
the distribution of the orientational order parameter $q$ (see Figure \ref{fig.struct}),
as defined by Errington \textit{et al.}
\cite{Debenedetti}: 
\be
q=1-\frac{3}{8}\sum_{j=1}^{3}\sum_{k=j+1}^{4}\biggl(cos\psi_{jk}+\frac{1}{3}\biggr)^2,
\label{Eq:q} 
\ee
where $\psi_{jk}$ is the angle formed by the lines
joining the oxygen atom of a given molecule and those of its nearest
neighbors $j$ and $k$. The parameter $q$ measures the extent to
which a molecule and its four nearest neighbors adopt a
tetrahedral arrangement.
The excellent agreement between the rdfs and the angular distribution 
suggests that the explicit H-bond directionality is responsible for
the local structure. On the contrary, 
the small deviation of the orientational order parameter distribution 
suggests that the electrostatic interactions are at least partly 
responsible for the overall tetrahedral arrangement of the system. 

\subsection*{Two-site interaction model}

A two-site simplified model of water changes the symmetry of the molecule 
by using two interacting sites instead of three, one corresponding to the oxygen
atom and the second one providing the directionality of the dipole moment. 
Since the dipole moment of TIP3P is
$\mu=2.35D$ and the charge separation is q=0.82e \cite{Jorgensen}, the
effective separation of the negative and positive charge centers is
$d=\frac{\mu}{q} \approx 0.5966$ \AA. Thus, we define the two-site model
by representing each water molecule by 
two spheres, which are joined by a rigid rod of length 0.5966 \AA. 
Two-site models have been previously proposed for water \cite{Vothjcp2005}, 
however existing models cannot exactly reproduce the radial structure 
of the reference all-atom water model.  
The two-site water model with parameters optimized by the procedure described above can reproduce remarkably well 
the rdfs for the different pairs of interactions: oxygen-oxygen with $f_{p}=4.5*10^{-5}$, oxygen-dipole with $f_{p}=1.4*10^{-4}$ and dipole-dipole with $f_{p}=3.2*10^{-5}$.
Figure \ref{fig.struct} shows the angular distribution and the orientational order parameter distribution obtained for the two-site model, in comparison
with the results of the all-atom and three-site models.
The excellent agreement between the rdfs suggests that the inclusion of a dipole center is sufficient for the correct radial structure. The deviation on the angular distribution from the three-site model indicates that the explicit H-bond directionality has a contribution to the angular structure (in addition to the electrostatic
contribution, as found in the three-site model).  

\subsection*{One-site interaction model}

In the one-site coarse-grained model, the water is represented by one
spherically symmetrical site having a mass $m_{O}+2m_{H}$. 
As demonstrated in Ref. \cite{water_jphys},
the one-site model can reproduce the center-of-mass rdf ($f_{p}=7.6*10^{-5}$) and thermodynamic 
properties of the reference all-atom water model with remarkable accuracy.
The optimized effective potential (see inset of Figure \ref{fig.hybrid}) has a
first primary minimum at about 2.8 \AA \; corresponding to the
first peak in the center-of-mass rdf. A second, slightly weaker
and significantly broader minimum at 4.5 \AA \; corresponds to the
second hydration shell. The combined effect of the two leads to
a local packing close to that of the all-atom TIP3P water. 
Our effective coarse-grained potential is quite different from the
previously suggested potentials\cite{Ashbaugh,Vothjcp2005,Gordon,Johnson}:
while in previous one-site models the deepest minimum
corresponds to the second hydration shell, the absolute minimum in
our model is found in the first shell.

The good structural agreement between the explicit and coarse-grained models
indicates that although our coarse-grained model of water is
spherically symmetric and therefore does not have any explicit
directionality, it approximately captures the correct local structure.
Furthermore, the angular and orientational order parameter distribution coincides with what obtained for the two-site model.
Since the one-site model can qualitatively predict the correct angular
distribution and is computationally the least expensive of all the coarse-grained models, it was our choice for the multiscale approach (presented in \cite{water_jphys}).\\
As a consequence of the reduced number of DOFs, there is a time
scale difference in the dynamics of the coarse-grained system with respect to the atomistic one.
The diffusion coefficient increases as the level of coarse-graining is increased, with the diffusion coefficient of the center of mass for the TIP3P model $D \approx 3.7*10^{-5} \frac{cm^2}{s}$, for the three-site model $D \approx 4.68*10^{-5} \frac{cm^2}{s}$,
for the two-site model $D \approx 7.54*10^{-5} \frac{cm^2}{s}$ and for
the one-site model $D \approx 8.1*10^{-5} \frac{cm^2}{s}$. 

Using the one-site coarse-grained model one can significantly speed
up the simulation, with a total gain in computational time of a factor
$\sim 17-20$ when compared to atomistic simulations\cite{water_jphys}. 
This is due to the reduction of the number of
interactions, which are also softer than in the atomistic case, and
due to an intrinsic time scale difference
in the diffusive dynamics of the coarse-grained system, that is
faster of about a factor 2 than the all-atom
simulations counterpart (see the above values for the diffusion constant).

\section{Matching the diffusive dynamics of the coarse-grained to the all-atom model}

\begin{figure}[h]
\centering
\includegraphics[height=\linewidth,clip=]{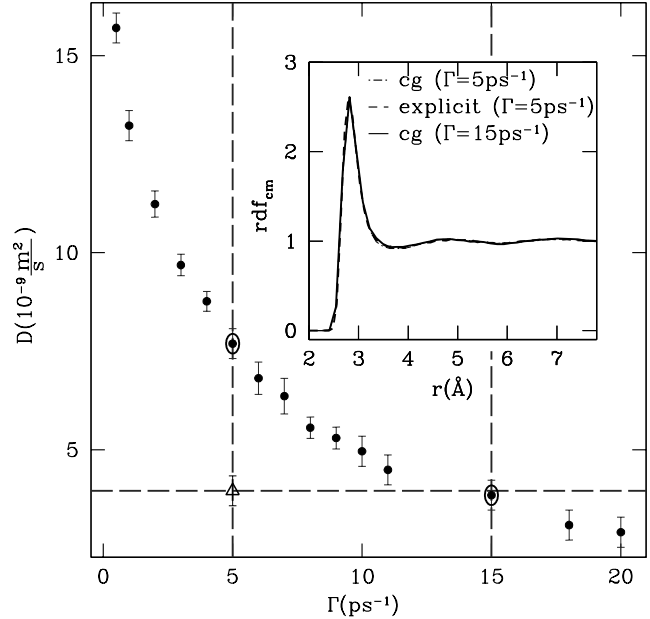}
\caption{\scriptsize 
The diffusion coefficient as a function of the friction constant.
The triangle indicates the value obtained for the all-atom diffusion coefficient when a value $\Gamma=5ps^{-1}$ is used in the simulation.
The black dots indicate the diffusion coefficient obtained for the
coarse-grained system for different values of $\Gamma$. The two
circles indicate the diffusion coefficients obtained for the
coarse-grained system when $\Gamma=5ps^{-1}$ and $\Gamma=15ps^{-1}$, respectively, are used in the simulation.
The inset compares the rdfs obtained for the all-atom system and the coarse-grained system for different values of the friction coefficients.    
\label{fig.diff} }
\end{figure}

As discussed above, the dynamics of the one-site coarse-grained model
is faster than that obtained from all-atom simulations. 
The speed-up occurs because of the reduction in DOFs upon coarse-graining, which eliminates the
fluctuating forces associated with those missing molecular DOFs,
leading to the much smoother overall energy landscape\cite{Tschop:1998,Izvekov:2006}. 
In our MD simulations we use a Langevin thermostat, and the equations of motion are in the form:

\be
m_{i}\frac{dv_{i}}{dt}=F_{i}-m_{i}\Gamma v_{i}+R_i(t)\label{langevin}
\ee
where $F_{i}=-\frac{\partial U_{i}}{\partial x_{i}}$ is the
deterministic force and $R_{i}$ is a stochastic variable with $\langle
R_{i}(t)R_{j}(t+\tau)\rangle=2\Gamma m_{i}kT\delta(\tau)\delta_{ij}$.
The coefficient $\Gamma$ determines the strength of the coupling to the bath [not to be confused 
with the friction coefficient $\xi$ of the system (for TIP3P
$\xi=288.6ps^{-1}$)]. When $\frac{1}{\Gamma}$ is large compared to the
typical time scales in the system, the stochastic dynamics reproduces
dynamics obtained by molecular dynamics (MD)\cite{Kremer:1990}, while if $\frac{1}{\Gamma}$ is small, then the stochastic dynamics deviates significantly from MD. 
Simulations performed using the the one-site model with same coupling
to the bath ($\Gamma=5ps^{-1}\ll\xi$) that is usually used for TIP3P
\cite{Tironi} produce an increase of a factor of about $2$ in the diffusion time
scale of the coarse-grained system. An accelerated dynamics can be
advantageous in some cases but as mentioned previously it can be a
problem if the dynamical properties themselves are under
investigation. The coarse-grained dynamics can be slowed down by
increasing the effective friction in the coarse-grained system. That
is, by changing  $\Gamma$, the one-site model can mimic the diffusive
dynamic of the all-atom system. Figure \ref{fig.diff} shows that the
diffusion coefficient D monotonically decreases as the coefficient
$\Gamma$ increases. The coarse-grained model yields the same diffusion
coefficient as the all-atom system (for which $\Gamma_{ex}=5ps^{-1}$)
when $\Gamma_{cg}=15 ps^{-1}$. The additional frictional noise does
not affect the short time dynamics of the 
system since $\Gamma_{ex}<\Gamma_{cg}\ll\xi$ \cite{Kremer:1990}. 
It is worth mentioning that the structural/thermodynamics of the 
coarse-grained system are not effected by 
changes in $\Gamma$ in this range. Figure \ref{fig.diff} shows the perfect agreement
between the center-of-mass rdf for the coarse-grained and all-atom
system when different values of $\Gamma$ are used for the two
models. However, we have to bear in mind that 
by employing the Langevin thermostat, Eq. (\ref{langevin}), we
screen the hydrodynamics. Furthermore, for too large $\Gamma$ the statics is also
altered because one arrives eventually at the Brownian dynamics. In order to
correctly reproduce the hydrodynamics one has to resort to the Dissipative
Particle Dynamics (DPD) thermostat\cite{Soddemann:2003} as was done
for instance in Ref.\cite{Praprotnik:2007:3} for the case of a
macromolecule in the hybrid atomistic/mesoscale solvent.

\section{Adaptive resolution scheme (AdResS)}

As we have proposed in \cite{water_jphys}, we design an adaptive 
multiscale system where half of the simulation box 
is occupied by atomistic water molecules and
the other half by the corresponding one-site
coarse-grained molecules, respectively, as schematically presented in Figure
\ref{fig.system}.
In order to smoothly couple the regimes of high and low level of detail, 
we apply the AdResS scheme \cite{Praprotnik,Praprotnik1,water_jphys},
that allows the molecules to move freely between regimes without feeling
any barrier in between.
The interface region contains
hybrid molecules that are composed of an all-atom molecule
with an additional massless center-of-mass particle serving as an
interaction site. The transition is governed by a weighting function $w$
that interpolates the molecular interaction forces between the two regimes,
and assigns the identity of the molecules.
In the present work we used the weighting function defined in
\cite{Praprotnik} and described in Figure \ref{fig.system}. 
The function $w$ is defined in such a way that $w=1$ corresponds to the
atomistic region, and $w=0$ to the coarse-grained region, whereas
the values $0<w<1$ correspond to the interface layer.
This interpolation leads to intermolecular forces acting on the center-of-mass of the molecules
$\alpha$ and
$\beta$ as: \be {\bf
F}_{\alpha\beta}=w(X_{\alpha})w(X_{\beta}){\bf
F}^{atom}_{\alpha\beta} + [1-w(X_{\alpha})w(X_{\beta})]{\bf
F}^{cm}_{\alpha\beta}.\label{eq:AdResS}\ee
${\bf F}_{\alpha\beta}$ is the total intermolecular force acting
between center of mass of molecules $\alpha$ and $\beta$.
${\bf F}^{atom}_{\alpha\beta}$ is the sum of all pair atomic interactions between
explicit water atoms of molecule $\alpha$ and explicit water atoms of molecule
$\beta$, and ${\bf F}^{cm}_{\alpha\beta}$ is the effective pair force between the center-of-mass of two water
molecules. $X_{\alpha}$ and $X_{\beta}$ are the center-of-mass coordinates of
molecules $\alpha$ and $\beta$. Note that the AdResS as given by
Eq. (\ref{eq:AdResS}) satisfies Newton's Third Law. This is crucial
for the diffusion of molecules across the resolution
boundaries\cite{Praprotnik:2007:2}. Since the total force as defined
by Eq. (\ref{eq:AdResS}) depends on the absolute positions of the
particles and not only on their relative distances it is in general not
conservative, i.e., the work done by it depends on the path taken in
the transition regime. Hence the corresponding potential does not
exist and the total potential energy of the system can not be
  defined\cite{Goldstein:1980, Ensing}. Still, the 
  intermolecular forces between molecules outside the transition
  regime are conservative with well defined potentials, i.e., the all-atom or effective
coarse-grained potentials\cite{Praprotnik}.

Each time a molecule crosses the
boundary between the different regimes it gains or looses
(depending on whether it leaves or enters the coarse-grained
region) its equilibrated rotational DOFs while retaining its
linear momentum. 
By this choice of interactions in the interface region, the hybrid molecule interacts with molecules
in the coarse-grained region on a coarse-grained level. On the other hand, the interactions of the hybrid molecules with the molecules in the explicit region are a combination of the explicit-explicit and cg-cg interactions to smoothly and efficiently equilibrate additional DOFs upon moving in the explicit regime \cite{Praprotnik,Praprotnik1}.
This change of resolution, which can be thought in terms of similarity
to a phase transition\cite{Praprotnik:2007}, requires to supply or remove latent heat and thus
must be employed together with i.e. a
Langevin thermostat \cite{Praprotnik}. 
The thermostat is coupled locally to the particle motion and provides a mean to
deliver or absorb the latent heat.

As in Ref. \cite{water_jphys}
the reaction field (RF) method is used, in which all molecules outside a spherical cavity of a molecular based
cutoff radius $R_c=9$ \AA \; are treated as a dielectric continuum
with a dielectric constant
$\epsilon_{RF}=80$\cite{Neumann:1983,Neumann:1985,Neumann:1986,Tironi,Im:2001,Praprotnik:2004}. 
The Coulomb force acting
on a partial charge $e_{i_{\alpha}}$, belonging to the explicit or hybrid molecule $\alpha$, at the center of the cutoff sphere,
due to a partial charge $e_{j_{\beta}}$, belonging to
the explicit or hybrid molecule $\beta$, within the cavity is: \be
 {\bf F}^{atom}_{C_{i_{\alpha}
j_{\beta}}}({\bf r}_{i_{\alpha} j_{\beta}})=
\frac{e_{i_{\alpha}}e_{j_{\beta}}}{4 \pi\epsilon_0}\biggl [\frac{1}{r_{i_{\alpha} j_{\beta}}^3}-\frac{1}{R_c^3}\frac{2(\epsilon_{RF}-1)}{1+2\epsilon_{RF}}
  \biggr]{\bf r}_{i_{\alpha} j_{\beta}}.\label{eq:rf}
\ee 

\begin{figure}[h]
\centering
\includegraphics[height=\linewidth,clip=]{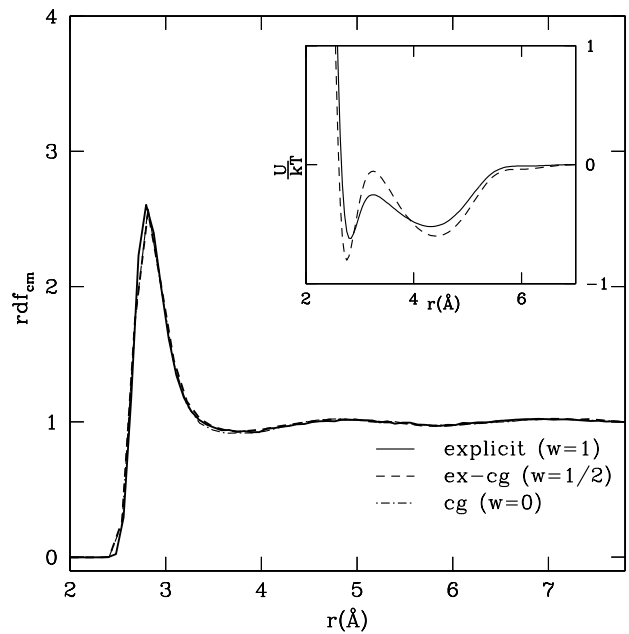}
\caption{\scriptsize The center-of-mass rdfs for explicit,
  ex-cg(w=1/2), and coarse-grained systems.
The inset shows the corrected effective pair potential $U^{cm}_{ic}$ for hybrid molecules [dashed line] and the
reference effective potential for the coarse-grained molecules [full line]. 
\label{fig.hybrid} }
\end{figure}
\begin{figure}[h]
\centering
\includegraphics[width=\linewidth,clip=]{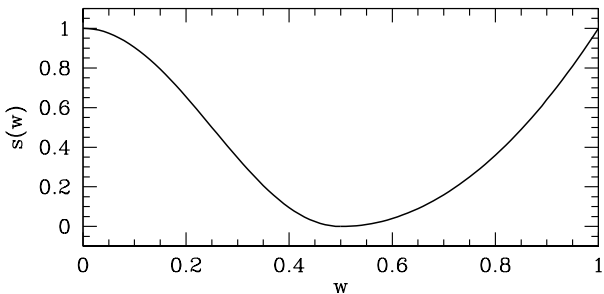}
\caption{\scriptsize The interface correction weighting function s(w). The value s=1 corresponds to the atomistic and coarse-grained regions and s=0 when w=1/2.
\label{s(w)}}
\end{figure}
\begin{figure}[h]
\centering
\includegraphics[height=\linewidth,clip=]{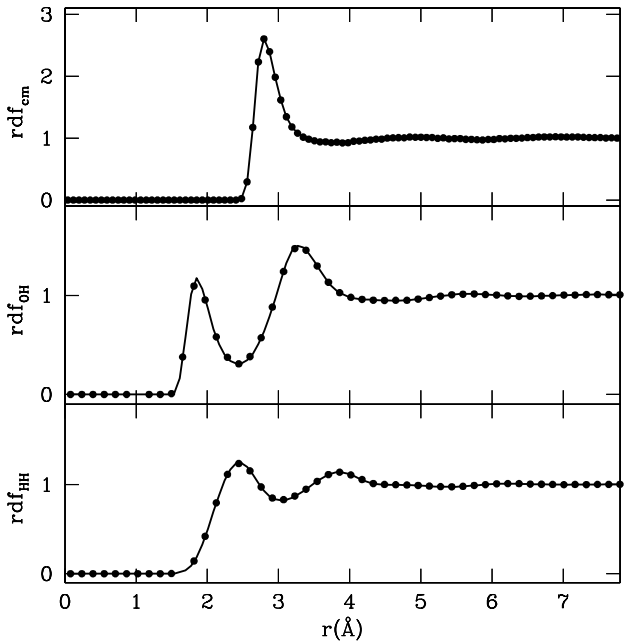}
\caption{\scriptsize The center-of-mass (fp=$7.65*10^{-5} $), OH (fp=$6*10^{-5}$) and HH (fp=$6.6*10^{-5}$) rdfs for the explicit region in the hybrid system [dots], and bulk [line] systems.
\label{fig_rdfs} }
\end{figure}

To suppress the unphysical pressure fluctuations emerging as artifacts
of the scheme given in Eq. \ref{eq:AdResS} we employ
an interface pressure correction \cite{Praprotnik1} within the transition zone.
The latter requires
a re-parametrization of the effective potential in the system composed of
exclusively hybrid molecules (with a constant of w=1/2).
We redefine the center-of-mass intermolecular forces
as
\be
{\bf F}^{cm}_{\alpha\beta}=
s[w(X_{\alpha})w(X_{\beta})]{\bf F}^{cm}_{\alpha\beta_{o}} +
\biggl(1-s[w(X_{\alpha})w(X_{\beta}])\biggr){\bf F}^{cm}_{\alpha\beta_{ic}},
\ee
where the function $s \in [0,1] $ is defined as
\begin{equation}
s[x]=
\begin{cases}
\cos(\pi\sqrt x)^{2}  & \mbox{   if  $0\leq x <0.25$} \\
4(\sqrt{x}-\frac{1}{2})^{2} & \mbox{if $0.25\leq x \leq 1.0$}
\end{cases}
\label{eq.s}
\end{equation}
so that s[0]=1, s[1]=1, and s[1/4]=0, as shown in Figure \ref{s(w)}.
The interface correction force $F^{cm}_{\alpha\beta_{ic}}$ is the effective pair force between
molecules $\alpha$ and $\beta$ defined by the effective pair potential, which is obtained
by mapping the hybrid model system composed of exclusively hybrid molecules with $w=1/2$
to the explicit model system. The corrected effective potential $U^{cm}_{ic}$ shown in Figure \ref{fig.hybrid} is obtained by mapping the $ex-cg(w=1/2)_{ic}$ system containing only hybrid molecules.
The minima of the effective potential $U^{cm}_{ic}$ become deeper, and a higher barrier separates the first and second hydration shell when compared to the center-of-mass effective potential $U^{cm}$.
The center-of-mass rdf of the hydrid system exactly reproduces the structure of the reference system as shown in Figure \ref{fig.hybrid}.\\
The mixing function in Eq. \ref{eq.s} can correctly reproduce the thermodynamic properties of
the ($ex-cg$) system with the same mean temperature (0.1 \% of difference) and
pressure (0.5 \% of difference).
Detailed comparisons between
the bulk explicit simulations and the explicit regime in our
hybrid setup prove that this approach does not alter the
structural properties of the water model studied.
In particular, Figure \ref{fig_rdfs} shows that the structural
properties of the explicit regime in the multiscale system are exactly the same as in bulk
explicit simulations.
No change is detected in the orientational preferences of water molecules near the interface region.
It is worth mentioning that similar results are obtained in the interface region (results not shown).

\begin{figure}[h]
\centering
\includegraphics[height=\linewidth,clip=]{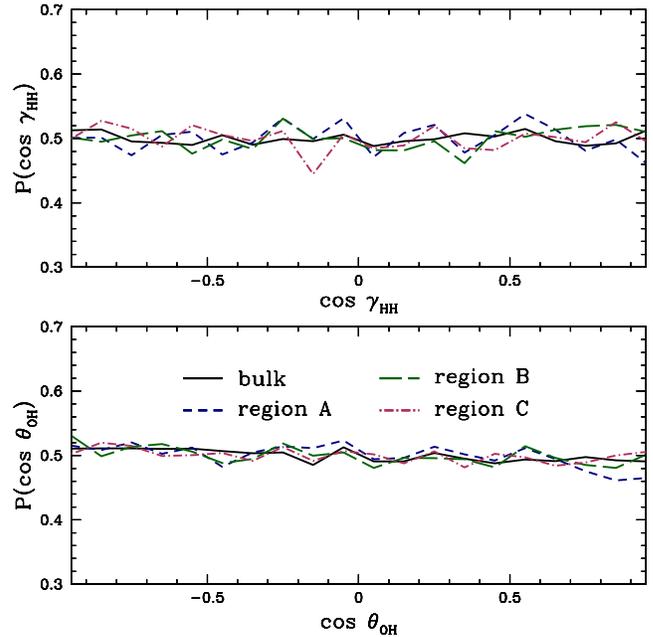}
\caption{\scriptsize (Color online) Cosine distribution of the angle formed by the H-H vector (top panel) and O-H vector (bottom panel) of the water molecules with the interface normal vector pointing toward the explicit region in three consecutives regions of width 4 \AA \; each (see Figure \ref{fig.system} for the definition of the regions). 
\label{fig_cosine} }
\end{figure}

Figure \ref{fig_cosine} shows the probability density function of the orientational parameters $\cos \theta_{OH}$ (O-H vector and the normal of the interface surface) and $\cos \gamma_{HH}$(H-H vector and the normal of the interface surface) in three consecutive layers of width 4 \AA \; next to the interface region along the X-axis (see Figure \ref{fig.system} for the definition of the regions).
The uniform cosine distribution of the water molecules in layers inside and next to the interface indicates that 
the orientational DOFs are fully equilibrated in the hybrid region, as
the molecules do not have to reorient upon crossing the interface region and there is no change of behaviour at the interface\cite{water_jphys}.

\section{Position dependent thermostat}
\begin{figure}[ht!]
\centering
\includegraphics[height=\linewidth,clip=]{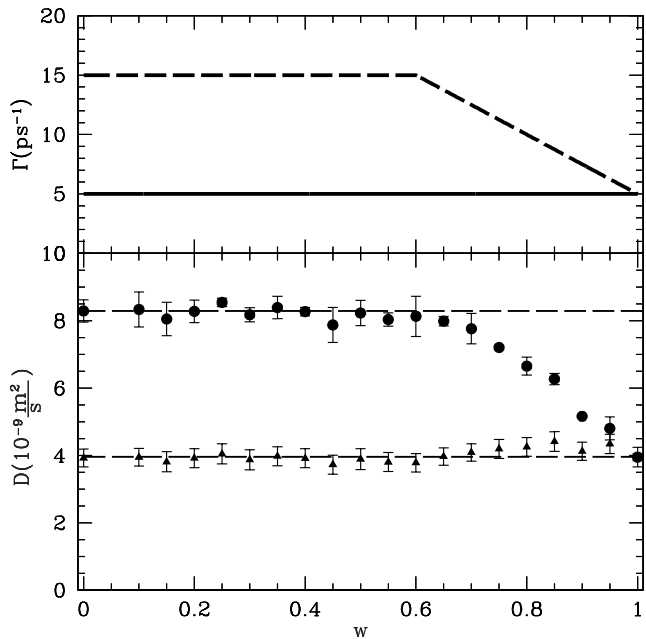}
\caption{\scriptsize 
Top  figure: The dashed curve indicates the dependency of the friction coefficient as a function of the particle identity $w$ when a position dependent thermostat is used. The full line shows the constant value of the friction coefficient when a regular thermostat is used.
Bottom figure: The dots indicate the diffusion of the molecules when the regular thermostat is used. The triangles indicate the diffusion of the molecules when the position dependent thermostat is used.
\label{fig.w_f} }
\end{figure}

\begin{figure}
\centering
\includegraphics[width=\linewidth,clip=]{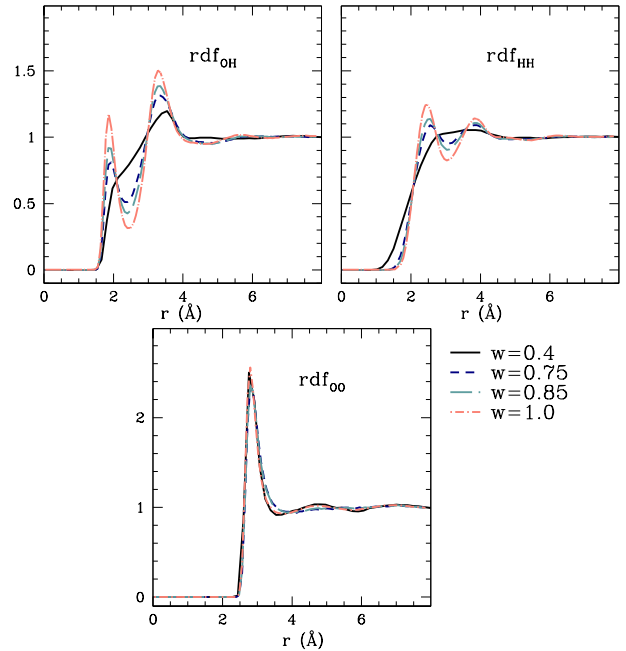}
\caption{\scriptsize (Color online) The H-H,O-H and O-O rdfs for different particles identities. The water molecules start to correct their structural properties when $w>0.6$. 
\label{fig.g}} 
\end{figure}

As mentioned above, the structural/thermodynamics of the coarse-grained system are not changed by increasing the background friction in the Langevin thermostat.
To obtain the same diffusional dynamics across different resolutions, the coefficient $\Gamma$ in the Langevin thermostat is changed on-the-fly depending on the number of DOFs of the molecules.
As shown in Figure \ref{fig.w_f}, when a constant coefficient $\Gamma$ is used, two regimes are observed in systems composed of only hybrid molecules:
\begin{itemize}
 \item For $w \; \epsilon \; [0-0.6]$ the value of the diffusion coefficient D is essentially constant; 
 \item For $w \; \epsilon \; ]0.6-1.0]$ the value of the diffusion coefficient D drops steeply with $w$.
 \end{itemize}
When the molecular identity $w$ is greater than $0.6$ the hybrid molecules start to equilibrate their orientational structure and their diffusive dynamics is slowed down, as shown in Figures \ref{fig.w_f} and \ref{fig.g}.
In order to obtain the same diffusive dynamics for different levels of resolution we propose the following functional form for the friction coefficient in the Langevin thermostat:
\begin{equation}
\Gamma(w)=
\begin{cases}
\Gamma_{cg}  & \mbox{   if  $w\leq0.6$} \\ 
\alpha w+\beta & \mbox{if $0.6< w \leq 1.0$} 
\end{cases}
\label{eq.gamma}
\end{equation}
This choice provides a simple interpolation between the two limit values of $\Gamma(0.6)=\Gamma(0)=\Gamma_{cg}=15ps^{-1}$ and $\Gamma(1)=\Gamma_{all-atom}=5ps^{-1}$. The parameters $\alpha$ and $\beta$ are 
$-25ps^{-1}$ and $30ps^{-1}$, respectively. As shown in Figure
\ref{fig.w_f} systems containing only hybrid molecules with different
particle identities exhibit the same diffusional dynamics when
$\Gamma(w)$ has the functional form proposed in Eq. \ref{eq.gamma}.

In our simulation we employ a local Langevin thermostat with a
position dependent friction \cite{carmeli,moix,grigolini,voth} to
match the diffusion constants of the coarse-grained and all-atom regimes.
The Langevin equation with a position dependent coefficient $\Gamma(x)$ can be written as \cite{carmeli}:
\be
m_{i}dv_{i}/dt=F_{i}-m_{i}\Gamma(x) v_{i}+R_i(x,t)
\ee
where $R_i(x,t)$ is:
\be
\langle R_i(x,t) \rangle =0, 
\ee
\be
\langle R_i(x,t_{1})R_j(x,t_{2}) \rangle=2\Gamma(x)m_{i}kT\delta(t_{1}-t_{2})\delta_{ij}
\ee
The functional form of $\Gamma(x)$ depends on the weighting function $w(x)$ and it is shown in Figure \ref{fig.w_f} (see Eq.\ref{eq.gamma}).
\begin{figure}[h]
\centering
\includegraphics[height=\linewidth,clip=]{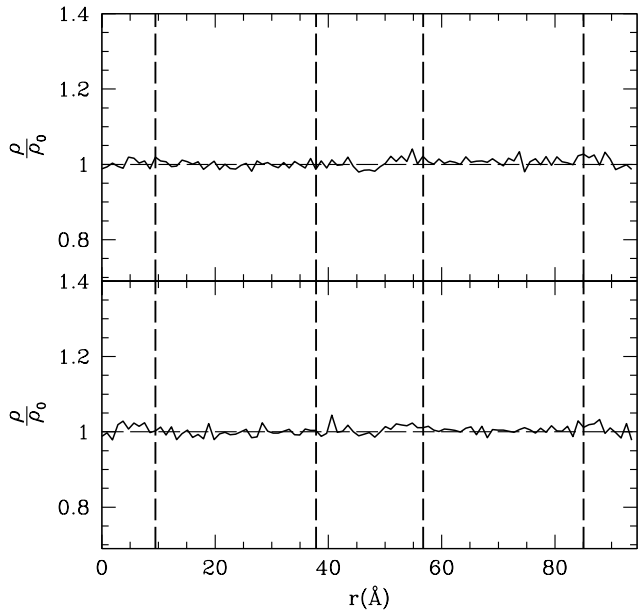}
\caption{\scriptsize Top figure: Normalized density profile in the x-direction
of the mixed system when the position dependent thermostat is used.
Bottom figure: Normalized density profile in the x-direction
of the mixed system when a constant friction constant is used.  
\label{fig.density} }
\end{figure}
There is no difference in the structural and thermodynamic properties
of the system when a position dependent $\Gamma(x)$ is used instead of a constant coefficient $\Gamma$. The density is homogeneous in both the coarse-grained and explicit regions with (very) small oscillations in the transition regime, cf. Figure \ref{fig.density} .\\
\begin{figure}[h]
\centering
\includegraphics[width=0.7\linewidth,clip=]{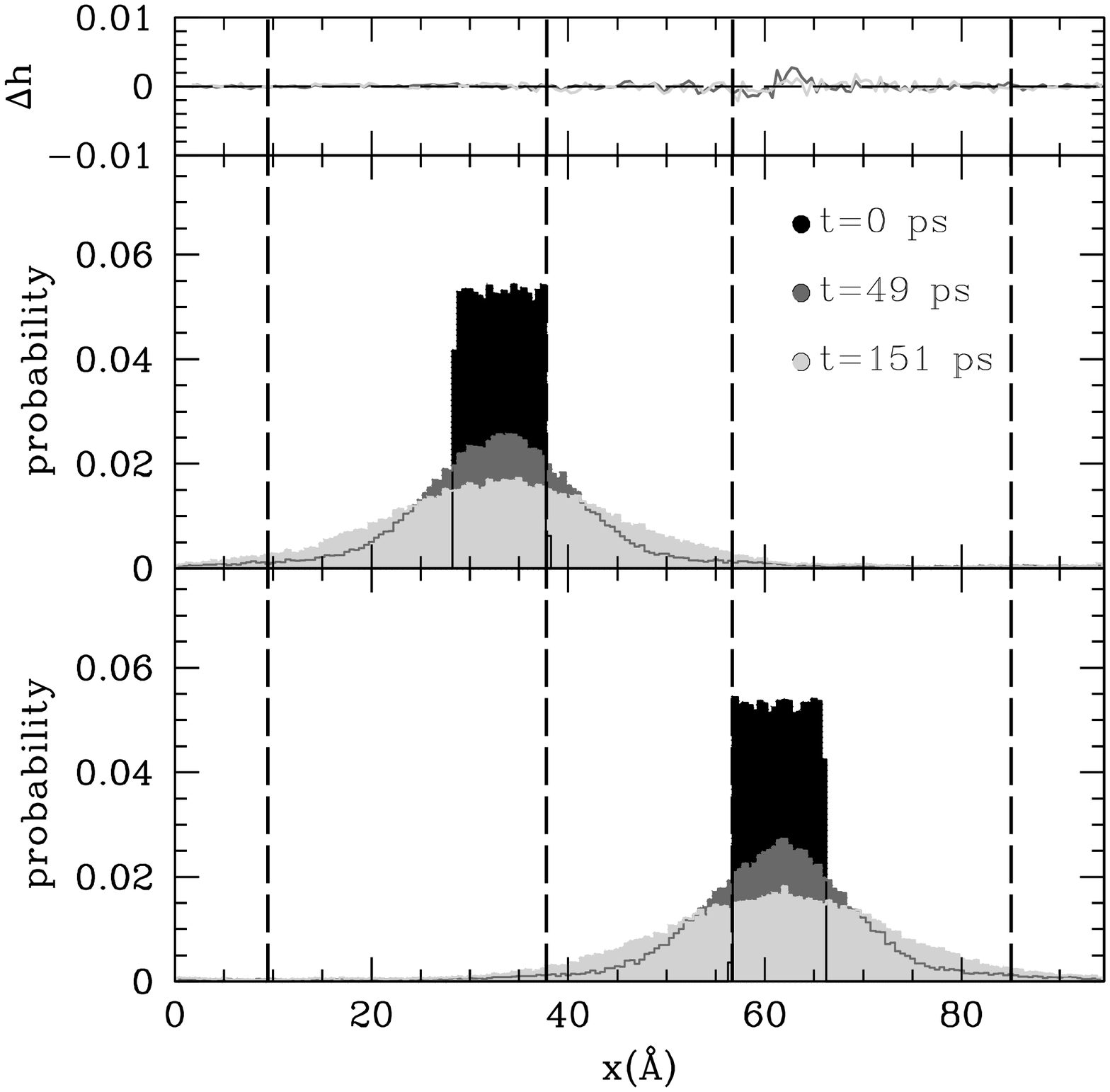}
\includegraphics[width=0.7\linewidth,clip=]{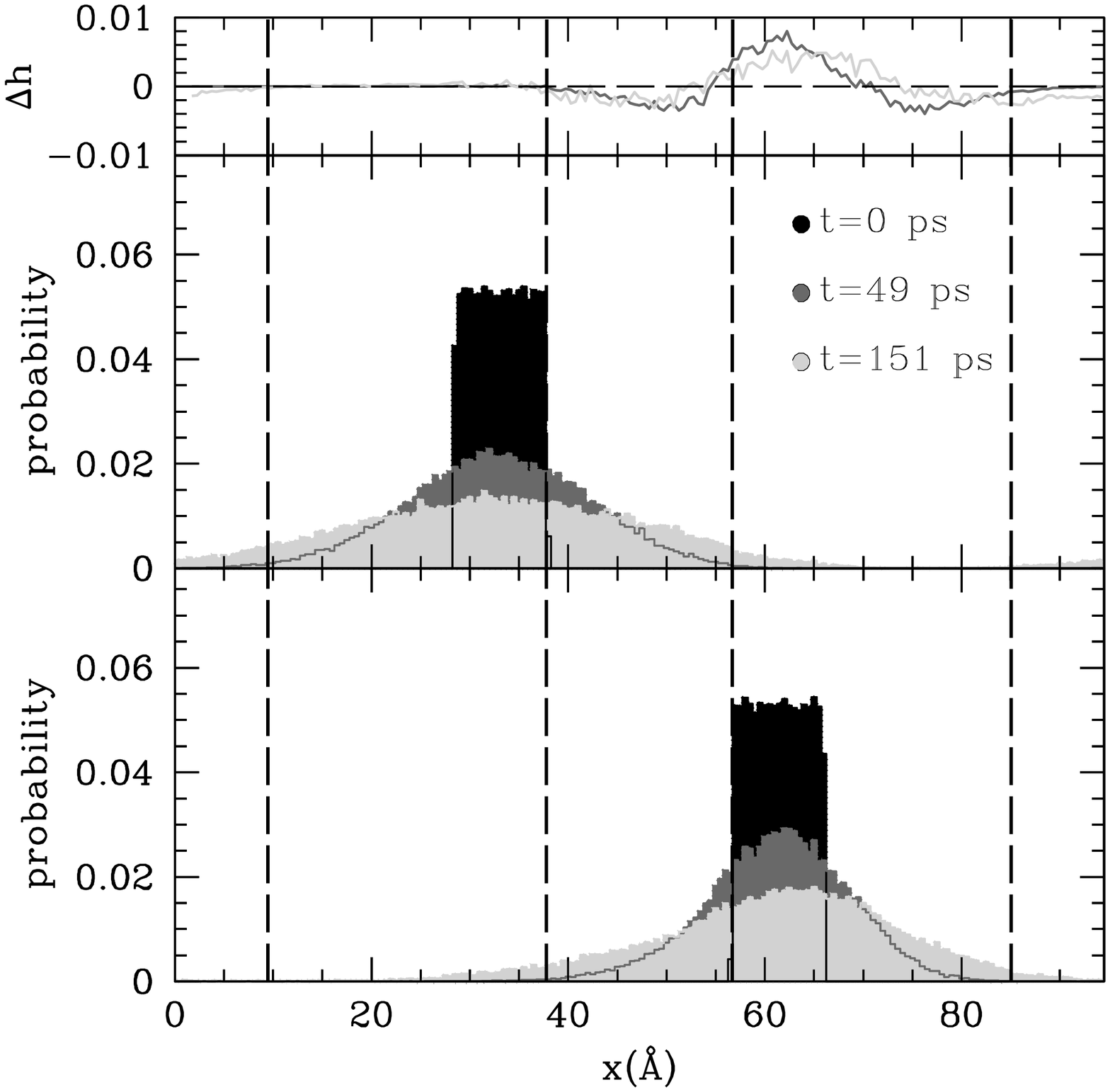}
\caption{\scriptsize (a) Time evolution of a diffusion profile when the position dependent thermostat is used for molecules that are initially (at time $t=0$ ps) located inside two neighboring slabs at opposite sides of the mid interface layer. The diffusion profile is averaged over $\approx 400$ different time origins.
Vertical lines denote the boundaries of the interface layer.
Bottom figure: The diffusion profile for the all-atom side is averaged over $\approx 400$ different time origins.
Middle figure: The same as before but for molecules that are initially localized inside the slab on the coarse-grained side of the interface region.
Top figure: Difference between the corresponding diffusion-profile of the coarse-grained and the all-atom regions.
(b) Time evolution of a diffusion profile with the regular thermostat using the same criteria as given in (a).  
\label{fig.histo} }
\end{figure}
To prove the free exchange of molecules between the different regimes we have computed the time evolution of a diffusion profile for molecules that were initially localized within two slabs of width $\approx 9.5$ \AA \; neighboring the interface layer.
The molecules initially localized within the two slabs spread out symmetrically with time when a position dependent friction is used in the local thermostat. On the contrary, the molecules spread out asymmetrically with time as shown in Figure \ref{fig.histo} when a fixed coefficient $\Gamma$ is used. This asymmetry arises from the above-mentioned difference in diffusion coefficient D between the all-atom and coarse-grained regions (see Ref. \cite{water_jphys}).
Figure \ref{fig.diff_comp} shows the behaviour of the diffusion coefficient D as a function of the x position using particles that are within a slab of $\approx 9.5$ \AA \;  of the mixed system for the two different type of thermostats.
As Figure \ref{fig.diff_comp} illustrates, the molecules are slowed down when they cross the interface from the coarse-grained region to the explicit region if a constant friction is used for the thermostat.  
The change in the diffusion coefficient D is localized in the interface region while D is approximately
constant inside each region, coarse-grained  [$D \approx (8.04\pm0.54) \times 10^{-9}
\frac{m^2}{s}$] and explicit [$D \approx (4.2\pm0.28) \times 10^{-9}
\frac{m^2}{s}$].
When a position dependent friction is used for the local thermostat
the diffusive dynamics of the molecules is the same as for the
all-atom system across all regions, as shown in Figure \ref{fig.diff_comp}.
\begin{figure}[h]
\centering
\includegraphics[height=\linewidth,clip=]{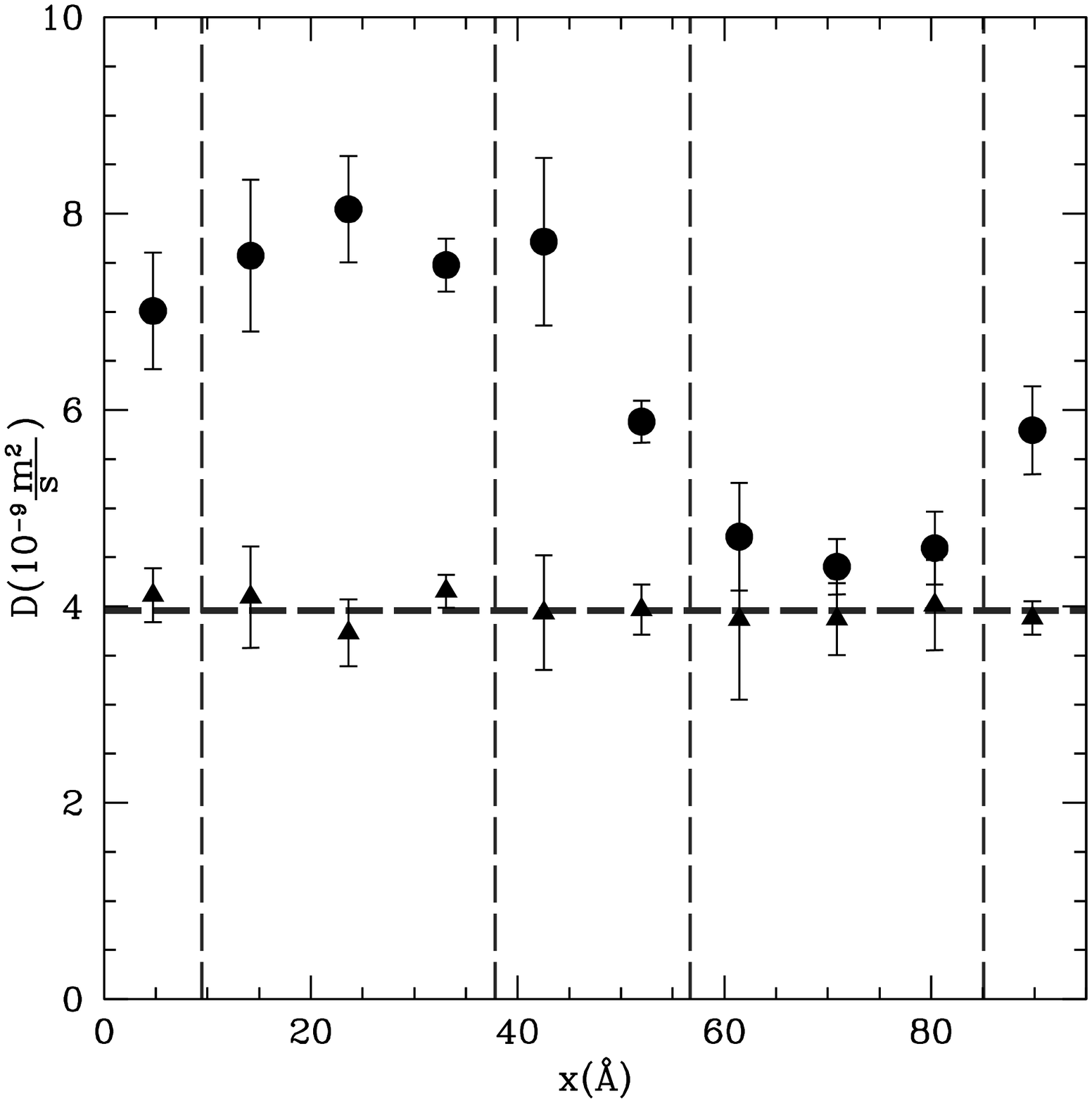}
\caption{\scriptsize 
Diffusion coefficient as a function of the x position using particles that are within a slab of $ \approx 9.5$ \AA \; of the hybrid system for different type of thermostats. Each point in the plot is centered in each slab. The average is over $\approx 400$ different time origins. The horizontal line is the diffusion coefficient of a system containing only all-atom water molecules.
The dots indicate the diffusion of the molecules across the hybrid system when the regular thermostat is used.
The triangles indicate the diffusion of the molecules across the hybrid system when the position dependent thermostat is used.
\label{fig.diff_comp} }
\end{figure}
\section{Conclusions}
In this paper we extended the multiscale model for
water we have recently proposed \cite{water_jphys} to obtain the same
diffusional dynamics across the different resolutions. 
 We also studied the accuracy of different coarse-grained water models 
in reproducing the structural properties of the all-atom system. 
  The results of this study show that for our purposes the single-site model performs equally well as the
  more sophisticated (three-site and two-site) coarse-grained models and is hence the most
  appropriate for the presented adaptive resolution simulations.
We envisage that such adaptive resolution simulations of water will play an important role in 
the modeling of biomolecular systems where the coupling of different time and length scales
is crucial to understand their physico-chemical properties.

\section*{Aknowledgments}
We wish to thank the NSF-funded Institute for Pure and Applied
Mathematics at UCLA where this work was first planned. This work
has been supported in part by grants from NSF, Texas-ATP, the
Robert A. Welch Foundation (C.C.) and the Volkswagen foundation
(K.K \& L.~D.~S.). The Rice Cray XD1 Cluster ADA used for the
calculations is supported by NSF, Intel, and Hewlett Packard.

\section{APPENDIX}
\subsection{Multiscale simulation protocol}

All the results presented in
the paper were obtained by performing nVT simulations using
the ESPResSo~\cite{espresso} (for the one-site and multi-scale model) and AMBER 8~\cite{AMBER} 
(for the 2-site and 3-site models) simulation packages with a Langevin thermostat, with a
friction constant $\Gamma=5$~ps$^{-1}$ when a regular thermostat is used 
and a time step of $0.002$~ps. All the models considered a system of of 1464
water molecules at $T_{ref}=300$~K and $\rho=0.96$~g/cm$^{3}$. The density was 
obtained from an all-atom NPT simulation with $P_{ref}=1$~atm using a
Reaction-field method for the electrostatics and a cut-off method for the
pressure.
The results presented for the 2-site and 3-site models were obtained for a
density $\rho=1.007$~g/cm$^{3}$. This value of the density was obtained from an
all-atom NPT simulation with $P_{ref}=1$~atm using an Ewald-method for the
electrostatics and a long-ranged correction for the pressure after the cut-off.
Periodic boundary conditions and minimum image convention were applied in all directions. 
The bonds and angle of the water molecules were constrained by using the RATTLE procedure. 
After warm-up and equilibration, several trajectories of $1.5$~ns each were
collected for each different model.
All simulations were performed with a force capping to prevent possible
force singularities that could emerge because of overlaps with neighboring molecules when
a given molecule enters the interface layer from the coarse-grained side.  The
size of the system is $94.5$~\AA \ in the x direction and $22$~\AA \ in both the y and z
directions. An interface layer of $18.9$~\AA \ between the coarse-grained and
all-atom models is set along the x direction. 
To treat all the molecules of the system equally regardless of their level of detail,
we define the pressure of the system according to their lower level of
detail, that is by the molecular pressure
\cite{Praprotnik,Praprotnik1, Berendsen:1984, Akkermans:2004}.
 The temperature was evaluated using the fractional analog of the
equipartition theorem:
\begin{equation}
 \left<K_\alpha\right>=\frac{\alpha k_BT}2,
\label{eq4}
\end{equation}
where $\left<K_\alpha\right>$ is the average kinetic energy per
fractional quadratic DOF with the weight
$w(r)=\alpha$\cite{Praprotnik:2007}. Via Eq. (\ref{eq4}) the
temperature is also rigorously defined in the transition regime in
which the rotational DOFs are partially 'switched
on/off'.

For each of the models considered, the following expression:
\begin{equation}
\begin{split}
\frac{V(r)}{kT}=&
a_0\biggl[\biggl(\frac{r_{0}}{r}\biggr)^{\alpha}-\biggl(\frac{r_{0}}{r}\biggr)^{\beta}\biggr] + 
a_1 e^{-b_1(r-r_{1})^{2}} +
a_2 e^{-b_2(r-r_{2})^{2}} \\ 
& +
a_3 e^{-b_3(r-r_{3})^{2}} +
a_4 e^{-b_4(r-r_{4})^{2}} +
a_5 e^{-b_5(r-r_{5})^{2}}
\end{split}
\label{eq_pot}
\end{equation}
is fitted within line thickness to the tabulated effective potential. 

\subsection{Three-site interaction model}
\begin{figure}[h!]
\centering
\includegraphics[width=\linewidth]{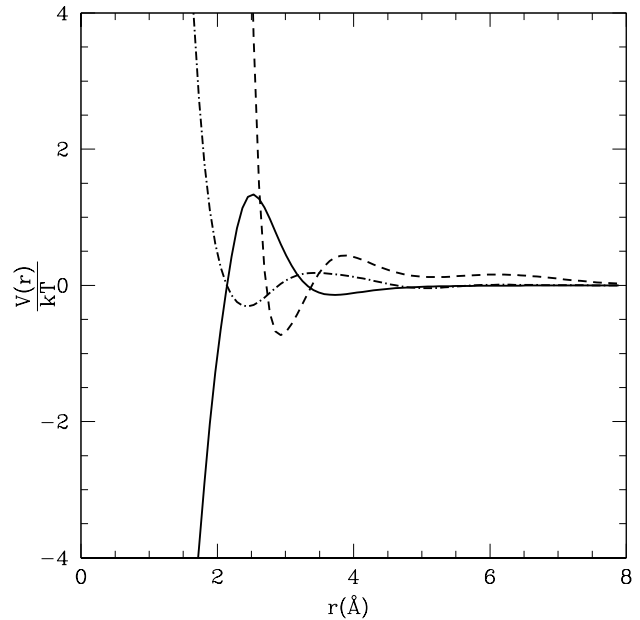}
\caption{\scriptsize
Effective pair potentials for the 3-site water model. The dashed,
dashed-dot, and full line curves indicate
the O-O, H-H, and O-H interactions, respectively.}
\label{fig.3site}
\end{figure}

\noindent
The coarse-graining procedure described in the paper converges after 12
iterations, yielding the effective potentials shown in Figure~\ref{fig.3site}. 

\subsection{Two-site interaction model}

\begin{figure}[h!]
\centering
\includegraphics[width=\linewidth]{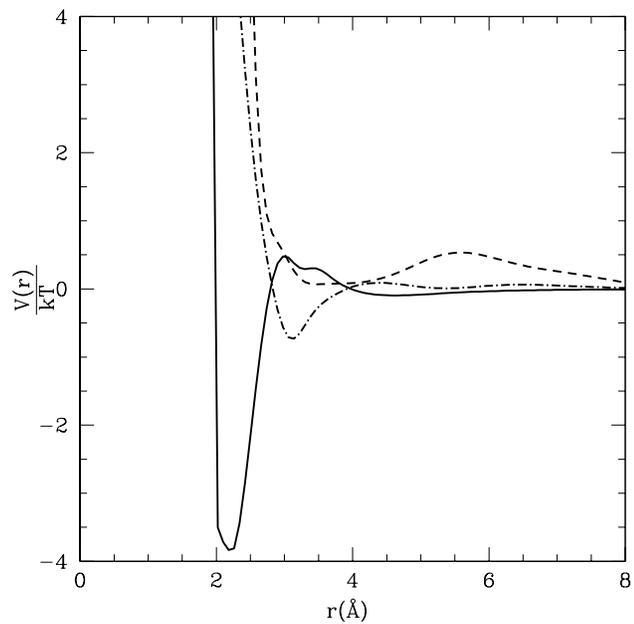}
\caption{\scriptsize
Effective pair potentials for the 2-site water model. The dashed,
dashed-dot, and full line curves indicate
the O-O, dipole-dipole, and O-dipole interactions, respectively.}
\label{fig.2site}
\end{figure}

\noindent
The coarse-graining procedure converges after 25 iterations; the resulting
effective potential is shown in Figure~\ref{fig.2site}. 

\subsection{One-site interaction model}

\begin{figure}[h!]
\centering
\includegraphics[width=\linewidth]{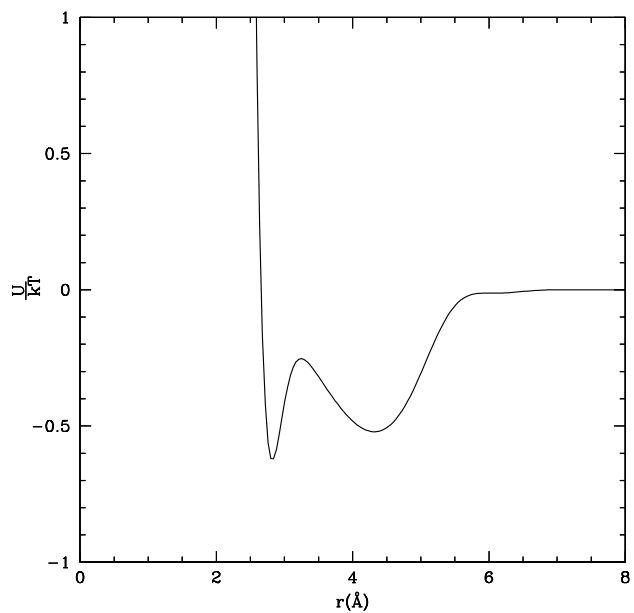}
\caption{\scriptsize
The effective interaction between the center-of-mass of two molecules as
obtained for the single-site water model.}
\label{fig.1site}
\end{figure}

\noindent
The coarse-graining procedure converges after 8 iterations.
The resulting potential for the effective interaction between the
center-of-mass of two molecules is shown in Figure~\ref{fig.1site}.





%
%

\end{document}